\documentstyle[prb,aps,twocolumn,epsf]{revtex}

\begin{document}

\twocolumn[\hsize\textwidth\columnwidth\hsize\csname
@twocolumnfalse\endcsname
\draft\title{Angle resolved Cu and O photoemission intensities 
in CuO$_{2}$ planes}
\author{J.M. Eroles, C.D. Batista and A. A. Aligia}
\address{Centro At\'{o}mico Bariloche and Instituto Balseiro,\\
Comisi\'on Nacional de Energ\'{\i}a At\'omica,\\
8400 Bariloche, Argentina}
\maketitle

\begin{abstract}
Using a mapping from the three-band extended Hubbard model for cuprate
superconductors into a generalized $t-J$ model, and exact diagonalization of
the latter in a 4$\times$4 cluster, we determine the quasiparticle weight
for destruction of a Cu or an O electrons with definite wave vector $k$. We
also derive an approximate but accurate analytical expression which relates
the O intensity with the quasiparticle weight in the generalized $t-J$ model.
The $k$-dependence of Cu and O intensities is markedly different. In
particular the O intensity vanishes for $k=(0,0)$. Our results are relevant
for the interpretation of angle-resolved photoemission experiments.
\end{abstract}
\vskip2pc]


\newpage

\section{Introduction}

While the wave vector dependence of the quasiparticle weight in quantum
antiferromagnets has been studied intensively since the discovery of high-$%
T_{c}$ superconductivity, the interest on the subject has been revived by
the angle-resolved photoemission (ARPES) experiments on insulating Sr$_{2}$%
CuO$_{2}$Cl$_{2}$ \cite{well}. Several theoretical works appeared fitting
the observed dispersion using generalized $t-J$ \cite
{naz,bel,xia,eder,lem1,leu,lem2}, spin-fermion \cite{sta} or one-band
Hubbard models \cite{esk,lem3}. More recently the photoemission intensities
have been discussed \cite{eder,lem1,leu,lem2,esk,sus} and compared with
previous results in the $t-J$ model \cite{szc,ede,poi}. In particular Lema
and one of us \cite{lem1,lem2} and Sushkov {\it et al.} \cite{sus}, have
developed two different methods to calculate the quasiparticle weight in the
generalized $t-J$ or one-band Hubbard models in the strong coupling limit,
using the self-consistent Born approximation (SCBA). The results of both
methods, an analytical approach based on the ''string picture'' \cite{ede},
and exact diagonalization of a 32-site cluster \cite{leu} were compared
recently \cite{lem2}. The method of Sushkov {\it et al.} introduces spurious
low-energy peaks in the Green function which can, however, be identified and
eliminated. The other SCBA method compares better with exact diagonalization
and the results of the string picture underestimate the weights. However,
since the operators $c_{k\sigma }^{\dag }$ entering generalized $t-J$ or
one-band Hubbard models are effective operators which cannot be trivially
translated into Cu and O holes of the original system, the above mentioned
efforts are insufficient for a comparison with experiment.

Experimental evidence about the symmetry of holes in cuprate superconductors 
\cite{nuc,tak,oda}, as well as constrained-density functional calculations 
\cite{ann,hyb}, justify the three-band Hubbard model $H_{3b}$ \cite{var,eme}
as the starting point for the description of these systems. $H_{3b}$
contains Cu 3d$_{x^{2}-y^{2}}$ and O 2p$_{\sigma }$ orbitals. To explain
some Raman \cite{ram1,ram2} and photoemission \cite{pot} experiments at
excitation energies above 1eV, it is necessary to include other orbitals in
the model \cite{sim}, but these are not important for the energy scale of
the ARPES experiments of Ref. 1. However, even restricting the basis to the
above mentioned orbitals, the size of the systems which can be diagonalized
numerically at present, do not contain more than four unit cells \cite{wag}.
This is too small to discuss the above mentioned ARPES experiments \cite
{well}. Thus, to study this problem by numerical methods at zero
temperature, it is necessary to integrate out the high-energy degrees of
freedom. Furthermore, analytical approximations like slave bosons give
better results when applied to an appropriate low-energy Hamiltonian \cite
{sim2}, and the successful SCBA cannot be applied to $H_{3b}$.

Several low-energy reduction procedures have been proposed \cite
{sim2,sch1,sf,sch2,sim3,fei1,bel2}. Eliminating the Cu-O hopping $t_{pd}$ by
means of a canonical transformation, leads to the spin-fermion $H_{sf}$ (or
Kondo-Heisenberg) model \cite{sch1,sf}. Although $t_{pd}$ is in principle
not small enough to guarantee the accuracy of the resulting $H_{sf}$, this
effective Hamiltonian, with parameters renormalized to fit the energy levels
of a CuO$_{4}$ cluster, reproduces very well optical and magnetic properties
of $H_{3b}$ in a Cu$_{4}$O$_{8}$ cluster \cite{sf}. Also one-band
generalized Hubbard \cite{sim2,sch1,sim3} and $t-J$ models \cite{fei1,bel2}
were derived. The latter represent the highest low-energy reduction reached
so far, and after the first proposal of Zhang and Rice \cite{zha}, further
work confirmed that a generalized $t-J$ model $H_{GtJ}$ reproduces
accurately the low-energy physics of the other models \cite
{hyb,ram,che,toh,bac,val,ali,ero}. In particular, projecting the Hilbert
space of $H_{sf}$ onto local (non-orthogonal) Zhang-Rice states \cite{ali},
mapping the model in this reduced Hilbert space to $H_{GtJ}$, and solving
numerically the latter, we obtained a band structure and magnetic properties
which agree very well with the corresponding properties calculated directly
on $H_{sf}$ \cite{ero}. It is important to emphasize that to calculate any
property of the cuprates, expressed in terms of the expectation value of an
operator of $H_{3b}$, using an effective low-energy {\em Hamiltonian}, the
mapping procedure should be extended to the {\em operator} of the quantity
to be calculated \cite{sf,sim3,fei}, or alternatively the relevant states of
the effective Hamiltonian should be mapped back onto the corresponding
states of $H_{3b}$.

In this work we calculate the low-energy part of the Cu and O ARPES, using
the low-energy reduction from $H_{3b}$ to $H_{sf}$ and from it to $H_{GtJ}$,
mapping the local Zhang-Rice singlets to vacant sites \cite{ali,ero}. The
relevant operators of $H_{3b}$ are mapped to $H_{sf}$, and the ground state
of $H_{sf}$ is constructed from that of $H_{GtJ}$ in a system containing 4$%
\times $4 unit cells. For the O ARPES we give a simple recipe to relate it
with the quasiparticle weight in $H_{GtJ}$, which can be calculated with the
SCBA or other analytical approaches \cite{lem2}. We find significant
differences between Cu and O ARPES. Since the photoemission cross-section
for Cu $d$ and O $p$ orbitals have different dependences on the incident
energy of the photon \cite{yeh}, these differences should be accessible to
experiments.

In section II we briefly review the mapping procedure and derive the
equations necessary to express the ARPES results in terms of numerical or
(in some cases) analytical results on $H_{GtJ}$. Section III contains the
results and section IV the conclusions.


\section{Mapping procedure and relevant equations}

\subsection{Deriving the spin-fermion model from $H_{3b}$}

Our starting point for the description of the superconducting cuprates below
1eV is the extended three-band Hubbard model. To simplify the writing we
change by -1 the phases of half of the O and Cu orbitals in such a way that
the hopping matrix elements do not depend on direction. The original phases
should be restored in the comparison with the experimental ARPES results 
\cite{note2}. The Hamiltonian takes the form:

\begin{eqnarray}
H_{3b} &=&\epsilon _{d}\sum_{i\sigma }d_{i\sigma }^{\dagger }d_{i\sigma
}+(\epsilon _{d}+\Delta )\sum_{j\sigma }p_{j\sigma }^{\dagger }p_{j\sigma }+
\nonumber \\
&&+U_{d}\sum_{i}d_{i\uparrow }^{\dagger }d_{i\uparrow }d_{i\downarrow
}^{\dagger }d_{i\downarrow }+U_{p}\sum_{j}p_{j\uparrow }^{\dagger
}p_{j\uparrow }p_{j\downarrow }^{\dagger }p_{j\downarrow }+  \nonumber \\
&&+U_{pd}\sum_{i\delta \sigma \sigma ^{\prime }}d_{i\sigma }^{\dagger
}d_{i\sigma }p_{i+\delta \sigma ^{\prime }}^{\dagger }p_{i+\delta \sigma
^{\prime }}  \nonumber \\
&&+t_{pd}\sum_{i\delta \sigma }\left[ p_{i+\delta \sigma }^{\dagger
}d_{i\sigma }+h.c.\right]  \nonumber \\
&&-t_{pp}\sum_{j\gamma \sigma }p_{j+\gamma \sigma }^{+}p_{j\sigma }
\label{1}
\end{eqnarray}

\noindent where $d_{i\sigma }^{\dagger }$ ($p_{j\sigma }^{\dagger }$)
creates a hole on the Cu 3d (O 2p) orbital at site $i$ ($j$). The four
nearest-neighbor (next nearest-neighbor) O sites to Cu site $i$ are denoted
by $i+\delta $ ($i+\gamma $). A canonical transformation which eliminates
terms linear in $t_{pd}$, retaining also the fourth order terms, leads to
the spin-fermion model \cite{sch1,sf}:

\begin{eqnarray}
H_{sf} &=&\sum_{i\delta \neq \delta ^{\prime }\sigma }\stackrel{\sim }{p}%
_{i+\delta ^{\prime }\sigma }^{\dag }\stackrel{\sim }{p}_{i+\delta \sigma
}[(t_{1}+t_{2})(\frac{1}{2}+2{\bf S}_{i}\cdot {\bf \Sigma }_{i+\delta
})-t_{2}]  \nonumber \\
&&-t_{pp}^{\prime }\sum_{j\gamma \sigma }\stackrel{\sim }{p}_{j+\gamma
\sigma }^{\dag }\stackrel{\sim }{p}_{j\sigma }+J_{K}\sum_{i\delta }({\bf S}%
_{i}\cdot {\bf \Sigma }_{i+\delta }-\frac{1}{4})  \nonumber \\
&&+\frac{J}{2}\sum_{i\delta }({\bf S}_{i}\cdot {\bf S}_{i+2\delta }-\frac{1}{%
4}),
\end{eqnarray}

\noindent where $\stackrel{\sim }{p}_{j\sigma }^{\dag }$ are effective O
creation operators, and ${\bf S}_{i}$ (${\bf \Sigma }_{i+\delta }$) is the
effective spin at Cu site $i$ (O site $i+\delta $). Due to the fact that $%
t_{pd}$ is not very small compared to $\Delta $ or $U_{d}-\Delta $, the
expressions for the parameters of $H_{sf}$ obtained from the canonical
transformation up to fourth order in $t_{pd}$ are not accurate enough.
However, this shortcoming is avoided if the parameters of $H_{sf}$ are
renormalized to fit the energy levels of $H_{3b}$ which in the limit $%
t_{pd}\rightarrow $ 0 corresponds to a level of $H_{sf}$, in a CuO$_{4}$
cluster with one and two holes. Since the case $t_{pp}\neq 0$ has not been
described before and the information is necessary for the expressions of the
ARPES results, we briefly review this method.

For two holes in the CuO$_{4}$ cluster the 16 eigenstates of $H_{sf}$ can be
classified in four spin singlets and four spin triplets distributed in six
energy levels: one $\Gamma _{1}$ (invariant under the point group
operations), one $\Gamma _{3}$ (transforming like $x^{2}-y^{2}$) and a
doublet $\Gamma _{5}$ (transforming like $x,y$) for each total spin. The
spin multiplicity 2S+1 will be denoted in the superscript. The ground state
is the invariant singlet $|g_{sf}\left( \Gamma _{1}^{1}\right) \rangle =%
\frac{1}{\sqrt{8}}\sum_{i\delta }(\tilde{p}_{i+\delta \uparrow }^{\dagger }%
\tilde{d}_{i\downarrow }^{\dagger }-\tilde{p}_{i+\delta \downarrow
}^{\dagger }\tilde{d}_{i\uparrow }^{\dagger })|0\rangle $, which represents
a Zhang-Rice singlet. Each eigenstate of $H_{sf}$ has a corresponding
eigenstate of $H_{3b}$, which is the lowest eigenstate of a small matrix in
the corresponding symmetry sector $\Gamma _{m}^{n}$. The largest matrix
corresponds to $\Gamma _{1}^{1}$ and is reproduced here for future use:

\begin{equation}
\left( 
\begin{array}{ccccc}
-2t_{pp} & 2t_{pd} & \sqrt{2}t_{pd} & \sqrt{2}t_{pd} & \sqrt{8}t_{pd} \\ 
2t_{pd} & \Delta & -2\sqrt{2}t_{pp} & -2\sqrt{2}t_{pp} & 0 \\ 
\sqrt{2}t_{pd} & -2\sqrt{2}t_{pp} & \Delta & 0 & 0 \\ 
\sqrt{2}t_{pd} & -2\sqrt{2}t_{pp} & 0 & \Delta +U_{p} & 0 \\ 
\sqrt{8}t_{pd} & 0 & 0 & 0 & U_{d}-\Delta -2U_{pd}
\end{array}
\right)  \label{matrix}
\end{equation}
The basis states of Eq.(\ref{matrix}) are the following:

\begin{eqnarray}
|1\rangle &=&\frac{1}{\sqrt{8}}\sum_{i\delta }(p_{i+\delta \uparrow
}^{\dagger }d_{i\downarrow }^{\dagger }-p_{i+\delta \downarrow }^{\dagger
}d_{i\uparrow }^{\dagger })|0\rangle ,  \nonumber \\
|2\rangle &=&\frac{1}{\sqrt{8}}\sum_{i\delta }(p_{i+\delta \uparrow
}^{\dagger }p_{i+R\delta \downarrow }^{\dagger }-p_{i+\delta \downarrow
}^{\dagger }p_{i+R\delta \uparrow }^{\dagger })|0\rangle ,  \nonumber \\
|3\rangle &=&\frac{1}{2}\sum_{i\delta }p_{i+\delta \uparrow }^{\dagger
}p_{i-\delta \downarrow }^{\dagger }|0\rangle ,  \nonumber \\
|4\rangle &=&\frac{1}{2}\sum_{i\delta }p_{i+\delta \uparrow }^{\dagger
}p_{i+\delta \downarrow }^{\dagger }|0\rangle ,  \nonumber \\
|5\rangle &=&d_{i\uparrow }^{\dagger }d_{i\downarrow }^{\dagger }|0\rangle ,
\label{basis}
\end{eqnarray}
where $R\delta $ is the result of rotating $\delta $ by $\pi /2$. To obtain
the optimum parameters of $H_{sf}$, we adjust them to fit exactly the three
lowest energy levels ($\Gamma _{1}^{1}$, $\Gamma _{5}^{1}$ and $\Gamma
_{5}^{3}$), the highest one ($\Gamma _{1}^{3}$), and the average of the
other two ($(\Gamma _{3}^{1}+\Gamma _{3}^{3})/2$). Calling $E_{3b}$ the
lowest energy of $H_{3b}$ in each symmetry sector, the result is: 
\begin{eqnarray}
t_{pp}^{\prime } &=&\frac{E_{3b}(\Gamma _{3}^{1})+E_{3b}(\Gamma
_{3}^{3})-E_{3b}(\Gamma _{5}^{1})-E_{3b}(\Gamma _{5}^{3})}{4}  \nonumber \\
t_{1} &=&\frac{E_{3b}(\Gamma _{1}^{3})-E_{3b}(\Gamma
_{5}^{3})+2t_{pp}^{\prime }}{4}  \nonumber \\
t_{2} &=&\frac{E_{3b}(\Gamma _{5}^{1})-E_{3b}(\Gamma _{1}^{1})}{8}-\frac{%
t_{1}}{2}-\frac{t_{pp}^{\prime }}{4}  \nonumber \\
J_{K} &=&2(t_{1}+t_{2})+E_{3b}(\Gamma _{5}^{3})-E_{3b}(\Gamma _{5}^{1})
\end{eqnarray}

\noindent As an example for the parameters of $H_{3b}$ for La$_{2}$CuO$_{4}$%
, obtained from constrained-density-functional approximation in Ref. \cite
{hyb} ($U_{d}=10.5,~U_{p}=4.0,~U_{pd}=1.2,~\Delta =3.6,~t_{pd}=1.3$ and $%
t_{pp}=0.6$, all energies in eV) , we obtain $t_{pp}^{\prime
}=0.56,~t_{1}=0.37,~t_{2}=0.08,~J_{K}=0.62$. The value of $J$, which is
affected by other orbitals not included in $H_{3b}$ \cite{bar} is taken as $%
J=0.13$ from experiment \cite{raman}.


\subsection{Mapping of the operators}

We have to express the hole creation operators $d_{i\sigma }^{\dagger }$ and 
$p_{i\sigma }^{\dagger }$ in the basis of $H_{sf}$ in order to calculate
photoemission properties. In the lowest non-trivial order in the canonical
transformation which eliminates $t_{pd}$, one obtains for the $d_{i\uparrow
}^{\dagger }$ operator transformed into the spin-fermion basis \cite{sf}: 
\begin{equation}
d_{i\uparrow }^{\dagger }=a\sum_{\delta }\widetilde{p}_{i+\delta \uparrow
}^{\dagger }\widetilde{n}_{i\uparrow }+b\sum_{\delta }\widetilde{p}%
_{i+\delta \uparrow }^{\dagger }\widetilde{n}_{i\downarrow }+c\sum_{\delta }%
\widetilde{p}_{i+\delta \downarrow }^{\dagger }\widetilde{d}_{i\uparrow
}^{\dagger }\widetilde{d}_{i\downarrow }  \label{d}
\end{equation}

\noindent and similarly interchanging spin up and spin down, with $%
\widetilde{n}_{i\sigma }=\widetilde{d}_{i\sigma }^{\dagger }\widetilde{d}%
_{i\sigma }$. The values of $a,~b,~c$ which are obtained from the canonical
transformation are not accurate enough. To improve them, we ask that all
matrix elements of the second member of Eq. (\ref{d}) between states of $%
H_{sf}$ in a CuO$_{4}$ cluster with one and two holes, should coincide with
the matrix elements of $d_{i\sigma }^{\dagger }$ between the corresponding
states in $H_{3b}$. The result is:

\begin{eqnarray}
a &=&v/2,\text{ }b=-u\left| A_{5}\right| /\sqrt{8}+\left( 1-\left|
A_{1}\right| \right) v/4  \nonumber \\
c &=&u\left| A_{5}\right| /\sqrt{8}+\left( 1+\left| A_{1}\right| \right) v/4,%
\text{ with }u,v>0  \nonumber \\
u^{2} &=&\frac{1}{2}+\frac{\Delta +U_{pd}-2t_{pp}}{2\sqrt{\left( \Delta
+U_{pd}-2t_{pp}\right) ^{2}+16t_{pd}^{2}}}  \nonumber \\
v^{2} &=&1-u^{2}  \label{coef,a,b,c}
\end{eqnarray}

\noindent The $A_{i}$ are the coefficients of the ground state of the matrix
Eq.(\ref{matrix}) in terms of the basis set Eq. (\ref{basis}): $%
|g_{3b}\left( \Gamma _{1}^{1}\right) \rangle =\sum\limits_{i}A_{i}|i\rangle $%
.

In the lowest non-trivial order in the canonical transformation, the
transformed operator of $p_{i\sigma }^{\dagger }$ is not changed. Following
a similar procedure as above, we assume that it is a good approximation to
use:

\begin{equation}
p_{i\sigma }^{\dagger }=a^{\prime }\widetilde{p}_{i\sigma }^{\dagger }
\label{ps}
\end{equation}

\noindent where $\left| a^{\prime }\right| <1$, because part of the spectral
weight of $p_{i\sigma }^{\dagger }$ is distributed in high-energy states
which are out of the Hilbert space of $H_{sf}$. Eqs.(\ref{ps}) and (\ref{d})
were shown to be accurate anough in previous comparison of the Cu and O
photoemission spectra of $H_{3b}$ and $H_{sf}$ in a Cu$_{4}$O$_{8}$ cluster 
\cite{sf}. Here, to calculate $a^{\prime }$, we solve exactly $H_{3b}$ and $%
H_{sf}$ in a Cu$_{2}$O cluster including and O atom and its two
nearest-neighbor Cu atoms, with two and three holes. The first (second)
member of Eq. (\ref{ps}) is applied to the S=0 ground state of $H_{3b}$ ($%
H_{sf}$) with two holes. For $H_{sf}$, the result is a linear combination of
two eigenstates with total spin S=1/2, which correspond to the low-energy
part of the result for $H_{3b}$. Then, $a^{\prime }$ is determined fitting
the coefficients of these two states. Inside the range of reasonable
parameters of $H_{3b}$, we obtain $\left| a^{\prime }\right| ^{2}\simeq
0.44. $


\subsection{From $H_{sf}$ to a generalized $t-J$ model}

There is numerical evidence\cite{val} that in the low-energy eigenstates of $%
H_{sf}$, the O holes are in the ground state of a CuO$_{4}$ cluster (a
Zhang-Rice singlet \cite{zha}) . Defined in this way, Zhang-Rice singlets
centered in nearest-neighbor Cu orbitals are non orthogonal \cite{note2}.
Using a projector $P_{2}$ over these non-orthogonal Zhang-Rice states \cite
{ali}, $P_{2}H_{sf}P_{2}$ can be mapped into a generalized $t-J$ model $%
H_{GtJ},$ in which each Zhang-Rice singlet at a CuO$_{4}$ cluster, is
replaced by the vacuum (no holes) in the cluster. Retaining the most
important terms, $H_{GtJ}$ takes the form \cite{ali}:

\begin{eqnarray}
H_{GtJ} &=&t_{1}^{\prime }\sum_{i\Delta \sigma }\tilde{d}_{i+\Delta \sigma
}^{\dagger }\tilde{d}_{i\sigma }+t_{2}^{\prime }\sum_{i\gamma \sigma }\tilde{%
d}_{i+\Gamma \sigma }^{\dagger }\tilde{d}_{i\sigma }  \nonumber \\
&&+t_{3}^{\prime }\sum_{i\Delta \sigma }\tilde{d}_{i+2\Delta \sigma
}^{\dagger }\tilde{d}_{i\sigma }  \nonumber \\
&&+t^{\prime \prime }\sum_{i\Delta \neq \Delta ^{\prime }\sigma }\tilde{d}%
_{i+\Delta ^{\prime }\sigma }^{\dagger }\tilde{d}_{i+\Delta \sigma }\left(
1-2{\bf S}_{i}\cdot {\bf S}_{i+\Delta }\right)  \nonumber \\
&&+\frac{J}{2}\sum_{i\Delta \sigma }\left( {\bf S}_{i}\cdot {\bf S}%
_{i+\Delta }-\frac{1}{4}\right)  \label{htj}
\end{eqnarray}

\noindent where $\Delta =2\delta $ ($\Gamma =2\gamma $) are vectors
connecting first (second) nearest-neighbor Cu atoms, and:

\begin{eqnarray}
t_{1}^{\prime } &=&\left( 104t_{pp}^{\prime
}+246t_{1}+410t_{2}+51J_{K}\right) /512  \nonumber \\
t_{2}^{\prime } &=&\left( 13t_{pp}^{\prime }-11t_{1}\right) /64  \nonumber \\
t_{3}^{\prime } &=&-11t_{1}/128  \nonumber \\
t^{\prime \prime } &=&\left( 8t_{pp}-18t_{1}-6t_{2}+3J_{k}\right) /256
\label{8}
\end{eqnarray}
As in $H_{sf}$ there is an implicit constrain of forbidden double occupancy
at any site. For the typical parameters of $H_{3b}$ mentioned above, Eqs. (%
\ref{8}) give: $t_{1}^{\prime }=0.42$, $t_{2}^{\prime }=0.05$, $%
t_{3}^{\prime }=0.06$, $t^{\prime \prime }=0.003.$

The low-energy eigenstates $|\Psi _{sf}^{\nu }\rangle $ of $H_{sf}$ can be
obtained from those $|\Psi _{GtJ}^{\nu }\rangle $ of $H_{GtJ}$ simply by 
{\it dressing} the vacant sites with Zhang-Rice singlets:

\begin{eqnarray}
|\Psi _{sf}^{\nu }\rangle &=&\frac{T|\Psi _{GtJ}^{\nu }\rangle }{\langle
\Psi _{GtJ}^{\nu }\left| T^{\dagger }T\right| \Psi _{GtJ}^{\nu }\rangle ^{%
\frac{1}{2}}}  \nonumber \\
\text{ }T &=&\prod\limits_{i}\left[ \frac{1}{\sqrt{8}}\sum_{\delta }\left( 
\stackrel{\sim }{p}_{i+\delta \uparrow }^{\dag }\tilde{d}_{i\downarrow
}^{\dagger }-\stackrel{\sim }{p}_{i+\delta \downarrow }^{\dag }\tilde{d}%
_{i\uparrow }^{\dagger }\right) \left( 1-n_{i}\right) +n_{i}\right] 
\nonumber \\
n_{i} &=&\tilde{d}_{i\uparrow }^{\dagger }\tilde{d}_{i\uparrow }+\tilde{d}%
_{i\downarrow }^{\dagger }\tilde{d}_{i\downarrow }  \label{transf}
\end{eqnarray}

\noindent This equation, together with Eqs.(\ref{d}) to (\ref{ps}) allow to
calculate the Cu and O photoemission spectra of the three-band Hubbard model 
$H_{3b}$ from the eigenstates of the corresponding generalized $t-J$ model $%
H_{GtJ}.$ For the sake of clarity, and in absence of a more detailed
knowledge about the experimental situation, we neglect effects of
interference and the dependence on the polarization of the incident
radiation and direction of the photoemitted electron. Then, in the
insulating state, the Cu and O contributions to the intensity of the lowest
ARPES peak are given by (the contributions for both spins are the same):

\begin{eqnarray}
I_{Cu\left( {\bf k}\right) } &=&2\left| \langle \Psi _{sf}^{{\bf k}}\left|
d_{{\bf k}\sigma }^{\dagger }\right| \Psi _{0}\rangle \right| ^{2}  \nonumber
\\
I_{O\left( {\bf k}\right) } &=&2\left| \langle \Psi _{sf}^{{\bf k}}\left| p_{%
{\bf k}x\sigma }^{\dagger }\right| \Psi _{0}\rangle \right| ^{2}+2\left|
\langle \Psi _{sf}^{{\bf k}}\left| p_{{\bf k}y\sigma }^{\dagger }\right|
\Psi _{0}\rangle \right| ^{2}  \label{inten}
\end{eqnarray}

\noindent where $|\Psi _{0}\rangle $ is the ground state of $H_{sf}$ and $%
H_{GtJ}$ in the insulating system, $|\Psi _{sf}^{{\bf k}}\rangle $ the
lowest energy eigenstate of $H_{sf}$ for one added hole (which leads to a
non-zero matrix element) and $d_{{\bf k}\sigma }^{\dagger }$, $p_{{\bf k}%
x\sigma }^{\dagger }$, and $p_{{\bf k}y\sigma }^{\dagger }$, are the Fourier
transforms of the three creation operators of a unit cell.


\subsection{O intensity vs quasiparticle weight in $H_{GtJ}$}

The formalism presented in the rest of this section allows us to calculate,
in the next section, the Cu and O contributions to ARPES, from the
eigenstates obtained from exact diagonalization of finite systems. To
calculate the Cu part with analytical approximations applied to $H_{GtJ}$
requires further algebraic elaboration which is beyond the scope of this
work. However, as we show below, there is a simple analytical relation
between the O contribution, generally the most important, and the
quasiparticle weight of $H_{GtJ}$ for one added hole. The latter quantity
has been calculated accurately with the SCBA \cite{lem1,lem2,sus} and
compared with results of other analytical and numerical methods \cite{lem2}.

The quasiparticle weight in $H_{GtJ}$ is:

\begin{equation}
Z_{\sigma }\left( {\bf k}\right) =\left| \langle \Psi _{GtJ}^{{\bf k}}\left| 
\tilde{d}_{{\bf k}\sigma }\right| \Psi _{0}\rangle \right| ^{2}
\label{def-Z}
\end{equation}

\noindent while the contribution to the intensity from, for example, 2p$_{x}$
orbitals and spin up is (Eqs.(\ref{ps}),(\ref{transf}), and (\ref{inten})):

\begin{eqnarray}
I_{O\sigma }^{x}\left( {\bf k}\right) &=&\left| a^{\prime }\right|
^{2}\left| \langle \Psi _{0}\left| \tilde{p}_{{\bf k}x\uparrow }\right| \Psi
_{sf}^{{\bf k}}\rangle \right| ^{2}  \nonumber \\
&=&\frac{\left| a^{\prime }\right| ^{2}\left| \langle \Psi _{0}\left| \frac{1%
}{\sqrt{N}}\sum_{j}^{^{\prime }}e^{-ikR_{j}}p_{j\uparrow }T\right| \Psi
_{GtJ}^{{\bf k}}\rangle \right| ^{2}}{N_{k}}  \label{11}
\end{eqnarray}

\noindent where $N_{k}=\left| \langle \Psi _{GtJ}^{{\bf k}}\left| T^{\dagger
}T\right| \Psi _{GtJ}^{{\bf k}}\rangle \right| ^{2}$ and in $%
\sum_{j}^{^{\prime }}$ the sun over $j$ runs over half the O atoms (those
which contains 2p$_{x}$ orbitals or in other words, their nearest-neighbor
Cu atoms lie in the $x$ direction). The norm $N_{k}\neq 1$, due to the
non-orthogonality of Zhang-Rice singlets centered in nearest-neighbor Cu
sites \cite{ali}. Since $|\Psi _{GtJ}^{{\bf k}}\rangle $ contains only one
vacant site then $(1-n_{i})(1-n_{j})|\Psi _{GtJ}^{{\bf k}}\rangle =0$ for $%
i\neq j$. Using this and Eq.(\ref{transf}) one has:

\begin{eqnarray}
\tilde{p}_{j}T|\Psi _{GtJ}^{{\bf k}}\rangle = &&\frac{1}{2\sqrt{2}}[\tilde{d}%
_{j+\delta _{x}\downarrow }^{\dagger }\left( 1-n_{j+\delta _{x}}\right)
\prod\limits_{i\neq j+\delta _{x}}n_{i}  \nonumber \\
&&+\tilde{d}_{j-\delta _{x}\downarrow }^{\dagger }\left( 1-n_{j-\delta
_{x}}\right) \prod\limits_{i\neq j-\delta _{x}}n_{i}]|\Psi _{GtJ}^{{\bf k}%
}\rangle  \nonumber \\
&=&\frac{1}{2\sqrt{2}}[\tilde{d}_{j+\delta _{x}\downarrow }^{\dagger }+%
\tilde{d}_{j-\delta _{x}\downarrow }^{\dagger }]|\Psi _{GtJ}^{{\bf k}}\rangle
\label{12}
\end{eqnarray}

\noindent where the last equality makes use of the fact that $n_{i}=0$ or 1
and $\sum\limits_{i}\left( 1-n_{i}\right) =1.$ From Eqs.(\ref{def-Z}) to (%
\ref{12}) one obtains:

\begin{equation}
I_{O\uparrow }^{x}\left( {\bf k}\right) =\frac{\cos ^{2}\left( \frac{k_{x}}{2%
}\right) \left| a^{\prime }\right| ^{2}}{2N_{k}}Z_{\downarrow }\left( -{\bf k%
}\right)  \label{13}
\end{equation}

\noindent and since from symmetry $Z_{\downarrow }\left( {\bf k}\right)
=Z_{\uparrow }\left( -{\bf k}\right) =Z_{\uparrow }\left( -{\bf k}\right) ,$
we have for the total O intensity \cite{note2}:

\begin{equation}
I_{O}\left( {\bf k}\right) =\frac{\left( \cos ^{2}\left( \frac{k_{x}}{2}%
\right) +\cos ^{2}\left( \frac{k_{y}}{2}\right) \right) \left| a^{\prime
}\right| ^{2}}{N_{{\bf k}}}Z_{\uparrow }\left( {\bf k}\right)  \label{14}
\end{equation}

\noindent In the cluster of 4$\times $4 unit cells, we obtain $N_{{\bf k}%
}\cong 0.36$ for all ${\bf k.}$ with error less than 10\%. The small
dependence of the norm $T|\Psi _{GtJ}^{{\bf k}}\rangle $ on wave vector is
to be expected in an antiferromagnetic background for realistic parameters
of $H_{GtJ}$. As it becomes particularly clear within the string picture 
\cite{ede}, the motion of a vacant site in a quantum antiferromagnet can be
divided into a fast motion around a fixed position on the lattice, on the
scale of $\sim 3t$, against a string linear potential created by the
distortion of the antiferromagnetic order, and a slow motion of the
polaronic cloud, which determines the quasiparticle dispersion (with a width 
$\sim 2J$). $N_{{\bf k}}$ is clearly determined by the physics inside the
polaronic cloud and is thus essentially independent of its wave vector ${\bf %
k.}$

From the above discussion, it is clear that the wave vector dependence of $%
I_{O}\left( {\bf k}\right) $ is given by that of the quasiparticle weight of 
$H_{GtJ},$ and the factor $\cos ^{2}\left( \frac{k_{x}}{2}\right) +\cos
^{2}\left( \frac{k_{y}}{2}\right) $\cite{note2}. This factor is very
important and leads to the fact that for wave vector $\left( \pi ,\pi
\right) $ in the notation of Eq. (\ref{1}) (${\bf k=}\left( 0,0\right) $
when the original phases are restored to compare with experiment \cite{note2}%
), there is no O contribution to the low-energy ARPES. This is particularly
clear when the on-site O repulsion $U_{p}=0$. In this case, from Eq. (\ref{1}%
), $\left[ H_{3b},p_{\left( \pi ,\pi \right) \alpha \sigma }^{\dagger
}\right] =\Delta p_{\left( \pi ,\pi \right) \alpha \sigma }^{\dagger }$ with 
$\alpha =x$ or $y$, {\it i.e.} $p_{\left( \pi ,\pi \right) \alpha \sigma
}^{\dagger }$ does not hybridize with the Cu 3$_{x^{2}-y^{2}}$ orbitals.
Then all the O weight resides in a well defined quasiparticle at energy $%
\Delta \sim 3.6$ eV, while the low-energy quasiparticles, involved in the
formation of Zhang-Rice singlets, lie at negative energies (with the zero of
one-particle energies of Eq.(\ref{1})\cite{sim}).


\section{Results}

In this section we present the result of exact diagonalization of $H_{GtJ}$
as an effective model representing the low-energy physics of $H_{3b}$, in a
system containing 4$\times $4 unit cells. At the end we use Eq.(\ref{14})
and previous results of $Z_{\uparrow }\left( {\bf k}\right)$ to obtain $%
I_{O}\left( {\bf k}\right) $ in larger clusters.

\begin{figure}
\narrowtext
\epsfxsize=3.5truein
\epsfysize=3.1truein
\vbox{\hskip 0.25truein \vskip 0.15truein
\epsfbox[3 20 582 750]{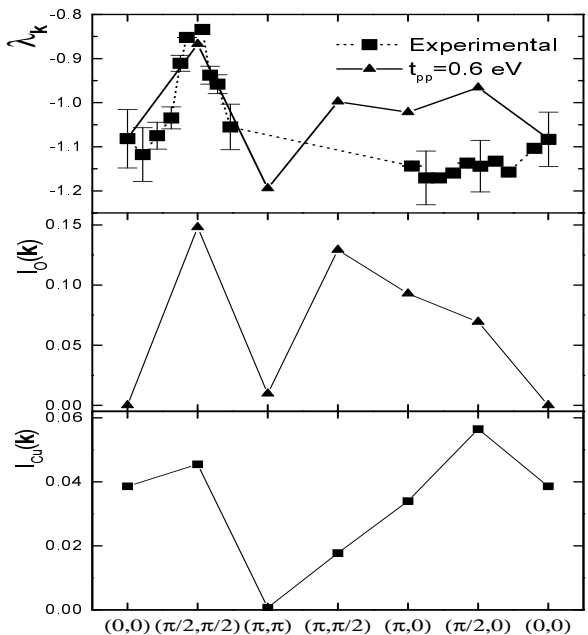}}
\medskip
\caption{Quasiparticle energies (top), oxygen intensity
(middle) and Cu intensity (bottom) as a function of wave vector, for
parameters of $H_{3b}$ calculated for La$_{2}$CuO$_{4}${\protect \cite{hyb}}. The
square symbols and error bars at the top correspond to the observed ARPES in
Sr$_{2}$CuO$_{2}$Cl$_{2}$ {\protect \cite{well}}.}
\end{figure}

At the top of Fig. 1 we show the quasiparticle dispersion $\lambda _{{\bf k}%
} $ with the original phases restored \cite{note2} and in the electron
representation (upside down with respect of the hole representation of $%
H_{3b}$), to facilitate comparison with experiment. The parameters of $%
H_{3b} $ were taken from Ref. \cite{hyb} and those of $H_{GtJ}$ were
determined from the mapping procedure, except for $J=0.13$ which was taken
from comparison with Raman experiments\cite{raman}. Taking into account that
there are no fitting parameters, the agreement with the experimentaly
measured dispersion in Sr$_{2}$CuO$_{2}$Cl$_{2}$ is very good. The
discrepancies around $\left( \pi ,0\right) $ can be ascribed to some
finite-size effects in the 4$\times $4 cluster \cite{poi}, and to the fact
that the parameters of $H_{3b}$ for La$_{2}$CuO$_{4}$ \cite{hyb} should
differ somewhat from the corresponding ones for Sr$_{2}$CuO$_{2}$Cl$_{2}.$ A
consequence of the upward shift in $\lambda _{{\bf k}}$ for ${\bf k=}\left(
\pi ,\frac{\pi }{2}\right) ,\left( \pi ,0\right) $ and $\left( \frac{\pi }{2}%
,0\right) $ is that the quasiparticle weight $Z_{\sigma }\left( {\bf k}%
\right) $ of $H_{GtJ}$ is exaggerated for these wave vectors \cite{lem2}.
This is due to the fact that for larger binding energy of the added hole,
less magnons are excited and the quasiparticle is more similar to the bare
hole, increasing $Z_{\sigma }\left( {\bf k}\right) $.

The O and Cu intensities given by Eqs.(\ref{inten}) are compared in Fig. 1.
As explained above, both intensities are exaggerated for wave vectors $%
\left( \pi ,\frac{\pi }{2}\right) ,\left( \pi ,0\right) $ and $\left( \frac{%
\pi }{2},0\right) .$ For $I_{O}\left( {\bf k}\right) $ this is clear when
Eq.(\ref{14})\cite{note2} with $\left| a^{\prime }\right| ^{2}/N_{{\bf k}%
}\cong 1.22,$ and the weights of $H_{GtJ}$ calculated by the SCBA \cite
{lem1,lem2} are used. However, these finite-size effects do not affect the
characteristic strong variation of the O intensity around the $\Sigma $ line
(joining $\left( 0,0\right) $ with $\left( \pi ,\pi \right) $). $I_{O}\left( 
{\bf k}\right) $ is maximum for ${\bf k}=\left( \frac{\pi }{2},\frac{\pi }{2}%
\right) $ and very small for ${\bf k}=\left( \pi ,\pi \right) ,$ as for the
generalized $t-J$ model \cite{lem1,leu,lem2,esk}. However, in contrast to $%
Z_{\sigma }\left( {\bf k}\right) $ for $H_{GtJ},$ $I_{O}\left( {\bf k}%
\right) $ vanishes at ${\bf k}=\left( 0,0\right) .$ This is a consequence of
the different symmetry of the $p_{\sigma }$ and d$_{x^{2}-y^{2}}$ orbitals
(or Zhang-Rice excitations) at that point, as explained (in different terms)
at the end of the previous section.

In contrast to $I_{O}\left( {\bf k}\right) ,$ the Cu intensities for ${\bf k}%
=\left( \frac{\pi }{2},\frac{\pi }{2}\right) $ and ${\bf k}=\left(
0,0\right) $ are similar and rather large in comparison with other wave
vectors. Since $I_{O}\left( 0,0\right) =0$, the experimental ARPES intensity
at ${\bf k=}\left( 0,0\right) $ is determined by the Cu part. The maximun of 
$I_{Cu}\left( {\bf k}\right) $ for ${\bf k}=\left( \frac{\pi }{2},0\right) $
is probably not realistic for the parameters of $H_{3b}$ which correspond to
Sr$_{2}$CuO$_{2}$Cl$_{2},$ and should be reduced as the corresponding $%
\lambda _{\frac{\pi }{2},0}$ approaches the observed quasiparticle energy.

\begin{figure}
\narrowtext
\epsfxsize=3.5truein
\epsfysize=3.1truein
\vbox{\hskip 0.25truein \vskip 0.15truein
\epsfbox[3 20 582 750]{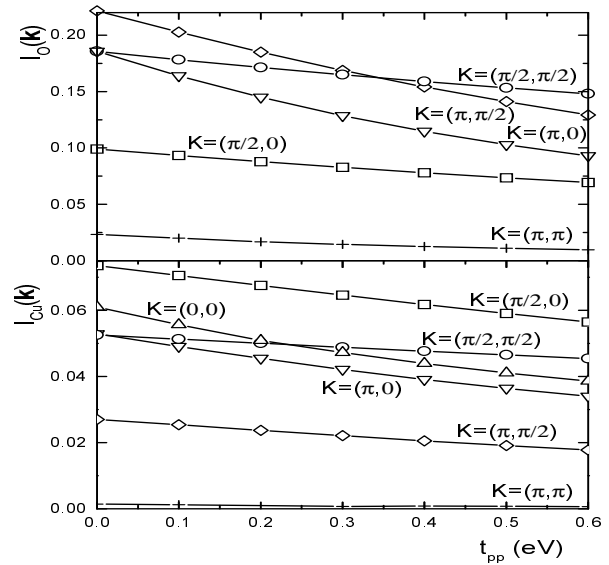}}
\medskip
\caption{Oxygen (top) and Cu (bottom) intensities for several
wave vectors as a function of O-O hopping.}
\end{figure}

In Fig. 2 we show the evolution of several intensities as a function of the
O-O hopping $t_{pp}$ of $H_{3b}.$ The importance of this term is that as $%
t_{pp}$ increases, the three site term $t^{\prime \prime }$ becomes positive
(see Eq.(\ref{8})), particularly if $t^{\prime \prime }$ is obtained by
fitting energy levels \cite{rvb1} instead of the analytical expression Eq. (%
\ref{8}) we used here. In turn, moderate positive values of $t^{\prime
\prime }$ favor a resonance-valence-bond superconducting ground state with
(predominantly) d$_{x^{2}-y^{2}}$ symmetry \cite{rvb1,rvb2}. The effect of $%
t_{pp}$ on the intensities is to reduce $I_{O}\left( {\bf k}\right) $ and $%
I_{Cu}\left( {\bf k}\right) $ for ${\bf k}=\left( \pi ,\frac{\pi }{2}\right)
,\left( \pi ,0\right) $ and $\left( \frac{\pi }{2},0\right) .$ Also $%
I_{Cu}\left( 0,0\right) $ decreases with $t_{pp}$. This is mainly a
consequence of a shift downwards of the corresponding $\lambda _{{\bf k}}$.
As a consequence, the dispersion, and also apparently the intensities,
compare better with the ARPES results in Sr$_{2}$CuO$_{2}$Cl$_{2}$, if $%
t_{pp}\sim 0.6$ eV or larger$.$

\begin{figure}
\narrowtext
\epsfxsize=3.5truein
\epsfysize=3.1truein
\vbox{\hskip 0.25truein \vskip 0.15truein
\epsfbox[3 20 582 750]{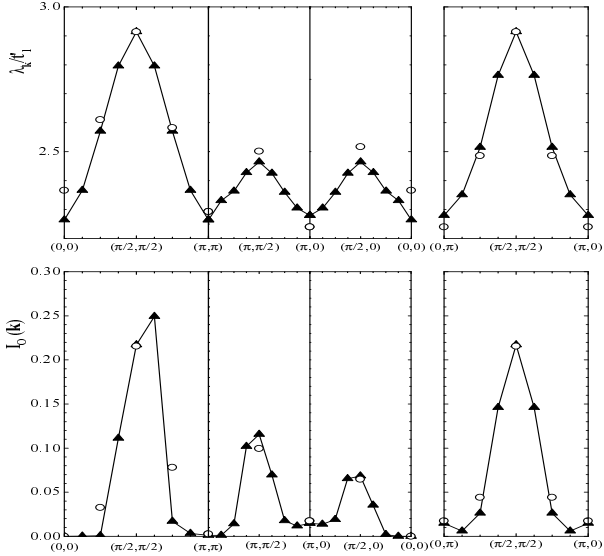}}
\medskip
\caption{Quasiparticle dispersion (top) and oxygen intensity (bottom) as a
function of wave vector using exact diagonalization (open circles) {\protect
\cite{leu}} and SCBA (solid triangles){\protect \cite{lem2}} results.
Parameters are $J=-t'_{2}=0.3t'_{1},~t'_{3}=0.2t'_{1}$ and $t''=0$.}
\end{figure}

For the incident energy used in the ARPES experiments \cite{well} in Sr$_{2}$%
CuO$_{2}$Cl$_{2}$ , the cross section for photoemiting O 2$p$ electrons is
near two times that of Cu 3$d$ electrons \cite{shen}. This fact and our
previous results suggest that the observed intensity is given essentially by 
$I_{O}\left( {\bf k}\right) $, except for ${\bf k}$ near $(0,0)$. We have
used Eq.(\ref{14}) with $N_{k}=0.36$ constant, in order to relate $%
I_{O}\left( {\bf k}\right) $ with previous accurate results for $Z_{\sigma
}\left( {\bf k}\right) $ in $H_{GtJ}$: exact diagonalization of a square
cluster of 32 sites \cite{leu}, and the SCBA in a 16$\times $16 cluster \cite
{lem2}. The parameters of $H_{GtJ}$, taken from Ref.\cite{leu}, are near the
optimum ones for fitting the dispersion relation $\lambda _{{\bf k}}$, with $%
t^{\prime \prime }=0$. The resulting $\lambda _{{\bf k}}$ and $I_{O}\left( 
{\bf k}\right) $ are shown in Fig. 3. Due to the factor $\sin ^{2}\left(
k_{x}/2\right) +\sin ^{2}\left( k_{y}/2\right) $ \cite{note2}, the
intensities along the $\Sigma $ line are asymmetric and smaller near the
Brillouin zone center, contrary to what was observed experimentaly. We
should state that a small admixture of the Cu 3$d^{10}$ configuration in the
ground state of the undoped system (of order $t_{pd}^{2}/[(\Delta
+U_{pd})U_{d}]$, which we have disregarded here) has the effect of
increasing the Cu ARPES near ${\bf k}=(0,0)$, as it is clear in the
strong-coupling limit of the one-band Hubbard model ($H_{1b}$)\cite
{lem1,esk,sus}. However, clearly this effect is negligible on the O ARPES.
The above mentioned asymmetry is even enlarged if a negative $t^{\prime
\prime }=-J/4$ (as that which comes form a canonical transformation of $%
H_{1b}$) is included \cite{leu,lem2}. This suggests again that values of $%
t_{pp}\sim 0.6$eV or larger, leading to positive $t^{\prime \prime }$, are
more realistic.

\section{Conclusions}

We have developed a formalism which allows to calculate separately the
low-energy part of the angle-resolved photoemission intensity from either O 2%
$p_{\sigma }$ or Cu 3d$_{x^{2}-y^{2}}$ orbitals, using a generalized $t-J$
model as an effective model for cuprate superconductors. This low-energy
reduction is the only way to calculate the wave vector dependence of the
intensities by exact diagonalization of finite systems, since at present it
is not possible to diagonalize directly the three-band Hamiltonian in a
periodic system large enough to contain the minimum necessary sampling of
the Brillouin zone.

For the insulating system, the O intensity can be very well approximated as 
\[
I_{O}\left( {\bf k}\right) \cong 1.22~Z_{\sigma }\left( {\bf k}\right)
\left( \sin ^{2}\left( k_{x}/2\right) +\sin ^{2}(k_{y}/2)\right) 
\]

\noindent where $Z_{\sigma }\left( {\bf k}\right) $ is the quasiparticle
weight of the effective generalized $t-J$ model. Thus, $I_{O}$ vanishes at
the $\Gamma $ point ${\bf k}=\left( 0,0\right) .$ Since this is a
consequence of the different symmetry of O 2$p_{\sigma }$ states and
low-energy excitations at that point, this result should persist with doping.

Our numerical results in a cluster of 4$\times $4 unit cells, for parameters
calculated for La$_{2}$CuO$_{4}$, show that $I_{O}\left( {\bf k}\right) $ is
largest for ${\bf k}=\left( \frac{\pi }{2},\frac{\pi }{2}\right) ,$ and at
that point, the Cu intensity $I_{Cu}\left( {\bf k}\right) $ is nearly three
times smaller. Instead while $I_{Cu}\left( {\bf k}\right) $ has similar
values at ${\bf k}=\left( \frac{\pi }{2},\frac{\pi }{2}\right) $ and near
the $\Gamma $ point, $I_{O}\left( 0,0\right) =0.$ The fact that $I_{O}\left( 
{\bf k}\right) $ and $I_{Cu}\left( {\bf k}\right) $ dominate in different
regions of the Brillouin zone, makes it possible to separate both
contributions experimentally. For an analysis of the experiments, as those
carried out in Sr$_{2}$CuO$_{2}$Cl$_{2}$\cite{well}, the separation in Cu
and O contributions is important, since the cross section for photoemiting
electrons in O 2$p$ or Cu 3d$_{x^{2}-y^{2}}$ orbitals are different and have
different dependence on the incident energy \cite{yeh}. For a quantitative
comparison with experiment, it is necessary add the amplitudes (instead of
the intensities) of the scattered waves from the three atoms per unit cell,
multiplied by their respective scattering amplitudes, taking into account
the polarization of the incident photons, and the direction of the
photoemited electrons. This does not require an extension of our formalism.
In addition, for any particular scattering amplitudes and polarization, the
expected trends can be extracted from the present results.

Agreement with the observed intensities seems to improve for $t_{pp}\geq0.6$%
eV, which in turn favors an RVB ground state and d-wave superconductivity 
\cite{rvb1,rvb2}.

\section*{Acknowledgments}

Two of us (JME and CDB) are supported by the Consejo Nacional de
Investigaciones Cient\'{\i }ficas y T\'{e}cnicas (CONICET), Argentina. (AAA)
is partially supported by CONICET.


\begin{references}
\bibitem{well}  B.O. Wells, Z.-X. Shen, A. Matsuura, D.M. King, M.A.
Kastner, M. Greven, and R.J. Birgeneau, Phys. Rev. Lett. {\bf 74,} 964
(1995).

\bibitem{naz}  A. Nazarenko, K.J.E. Vos, S. Haas, E. Dagotto, and R.
Gooding, Phys. Rev. B {\bf 51}, 8676 (1995).

\bibitem{bel}  V.I. Belinicher, A.L. Chernyshev, and V.A. Shubin, Phys. Rev.
B {\bf 54}, 14914 (1996).

\bibitem{xia}  T. Xiang and J.M. Wheatley, Phys. Rev. B {\bf 54}, R12653
(1996).

\bibitem{eder}  R. Eder, Y. Ohta, and G. A. Sawatzky, Phys. Rev. B {\bf 55},
R3414 (1997).

\bibitem{lem1}  F. Lema and A.A. Aligia, Phys. Rev. B {\bf 55}, 14092 (1997).

\bibitem{leu}  P.W. Leung, B.O. Wells and R.J. Gooding, Phys. Rev. B {\bf 56}%
, 6320 (1997); references therein.

\bibitem{lem2}  F. Lema and A.A. Aligia, Physica C {\bf 307}, 307 (1998).

\bibitem{sta}  O.A. Starykh, O.F. de Alcantara Bonfim, and G. Reiter, Phys.
Rev. B {\bf 52}, 12534 (1995).

\bibitem{esk}  H. Eskes and R. Eder, Phys. Rev. B {\bf 54}, 14226 (1996).

\bibitem{lem3}  F. Lema, J. Eroles, C.D. Batista and E. Gagliano, Phys. Rev.
B {\bf 55}, 15289 (1997).

\bibitem{szc}  K. v. Szcepanski, P. Horsch, W. Stephan, and M. Ziegler,
Phys. Rev. B {\bf 41}, 2017 (1990); V. Elser, D.A. Huse, B.I. Shraiman, and
E.D. Siggia, {\it ibid } {\bf 41}, 6715 (1990); E. Dagotto, R. Joynt, A.
Moreo, S. Bacci, and E. Gagliano, {\it ibid } {\bf 41}, 9049 (1990).

\bibitem{ede}  R. Eder and K.W. Becker, Phys. Rev. B {\bf 44}, 6982 (1991).

\bibitem{poi}  D. Poilblanc, T. Ziman, H.J. Schulz, and E. Dagotto, Phys.
Rev. B {\bf 47}, 14267 (1993).

\bibitem{sus}  O.P. Sushkov, G.A. Sawatzky, R. Eder, and H. Eskes, Phys.
Rev. B {\bf 56}, 11769 (1997).

\bibitem{nuc}  N. N\"{u}cker, H. Romberg, X.X. Xi, J. Fink, Gebenheimer and
Z.X. Zhao, Phys. Rev. B {\bf 39}, 6619 (1989).

\bibitem{tak}  M. Takigawa, P.C. Hammel, R.H. Heffner, Z. Fisk, K.C. Ott,
and J.D. Thomson, Phys. Rev. Lett. {\bf 63,} 1865 (1989).

\bibitem{oda}  M. Oda, C. Manabe, and M. Ido, Phys. Rev. B {\bf 53}, 2253
(1996).

\bibitem{ann}  J.F. Annet, R.M. Martin, A.K. McMahan, and S. Satpathy, Phys.
Rev. B {\bf 40}, 2620 (1989).

\bibitem{hyb}  M.S. Hybertsen, E.B. Stechel, M. Schl\"{u}ter, and D.R.
Jennison, Phys. Rev. B {\bf 41}, 11068 (1990); references therein.

\bibitem{var}  C.M. Varma, S. Schmitt-Rink, and E. Abrahams, Solid State
Commun. {\bf 62}, 681 (1987).

\bibitem{eme}  V.J. Emery, Phys. Rev. Lett. {\bf 58,} 2794 (1987).

\bibitem{ram1}  R. Liu, D. Salamon, M. Klein, S. Cooper, W. Lee, S.-W.
Cheong, and D. Ginsberg, Phys. Rev. Lett. {\bf 71,} 3709 (1993).

\bibitem{ram2}  D. Salamon, R. Liu, M. Klein, M. Karlow, S. Cooper, S.-W.
Cheong, W. Lee, and D. Ginsberg, Phys. Rev. B {\bf 51}, 6617 (1995).

\bibitem{pot}  J.M. Pothuizen, R. Eder, M. Matoba, G. Sawatzky, N.T. Hien,
and A.A. Menovsky, Phys. Rev. Lett. {\bf 78,} 717 (1997).

\bibitem{sim}  M.E. Simon, A.A. Aligia, C.D. Batista, E.R. Gagliano, and F.
Lema, Phys. Rev. B {\bf 54}, R3780 (1996).

\bibitem{wag}  J. Wagner, W. Hanke, and D.J. Scalapino, Phys. Rev. B {\bf 43}%
, 10517 (1991).

\bibitem{sim2}  M.E. Simon and A.A. Aligia, Phys. Rev. B {\bf 48}, 7471
(1993); references therein.

\bibitem{sch1}  H.-B. Sch\"{u}ttler and A.J. Fedro, J. Appl. Phys. {\bf 63},
4209 (1988); P. Prelov\v {s}ek, Phys. Lett. A {\bf 126}, 287 (1988); J.
Zaanen and A.M. Ole\'{s}, Phys. Rev. B {\bf 37}, 9423 (1988); M. Matsukawa
and H. Fukuyama, J. Phys. Soc. Jpn. {\bf 58}, 2845 (1989).

\bibitem{sf}  C.D. Batista and A.A. Aligia, Phys. Rev. B {\bf 47}, 8929
(1993); references therein.

\bibitem{sch2}  H.-B. Sch\"{u}ttler and A.J. Fedro, Phys. Rev. B {\bf 45},
7588 (1992).

\bibitem{sim3}  M.E. Simon, A.A. Aligia and E.R. Gagliano, Phys. Rev. B {\bf %
56, }5637 (1997); references therein.

\bibitem{fei1}  L.F. Feiner, J.H. Jefferson, and R. Raimondi, Phys. Rev. B 
{\bf 53,} 8751 (1996); references therein.

\bibitem{bel2}  V.I. Belinicher, A.L. Chernyshev, and L.V. Popovich, Phys.
Rev. B {\bf 50}, 13768 (1994); references therein.

\bibitem{zha}  F.C. Zhang and T.M. Rice, Phys. Rev. B {\bf 37}, 3759 (1988).

\bibitem{ram}  A. Ramsak and P. Prelov\v {s}ek, Phys. Rev. B {\bf 40}, 2239
(1989).

\bibitem{che}  C.-X. Chen, H.-B. Sch\"{u}ttler and A.J. Fedro, Phys. Rev. B 
{\bf 41}, 2581 (1990).

\bibitem{toh}  T. Tohyama and S. Maekawa, J. Phys. Soc. Jpn. {\bf 59}, 1760
(1990).

\bibitem{bac}  S.B. Bacci, E. Gagliano, R. Martin, and J. Annet, Phys. Rev.
B {\bf 44}, 7504 (1991).

\bibitem{val}  C.D. Batista and A.A. Aligia, Phys. Rev. B {\bf 48}, 4212
(1993); {\bf 49}, 6436(E) (1994).

\bibitem{ali}  A.A. Aligia, M.E. Simon, and C.D. Batista, Phys. Rev. B {\bf %
49}, 13061 (1994); C.D. Batista and A.A. Aligia, Phys. Rev. B {\bf 49},
16048 (1994).

\bibitem{ero}  J. Eroles, C.D. Batista and A.A. Aligia, Physica C {\bf 261,}
237 (1996).

\bibitem{fei}  L. Feiner, Phys. Rev. B {\bf 48}, 16857 (1993).

\bibitem{yeh}  J.J. Yeh and I. Lindau, Atomic Data and Nuclear Data Tables 
{\bf 32}, 1 (1985).

\bibitem{note2}  The change of phases by -1 in half of the O and Cu
orbitals, simplifies the notations of $H_{3b}$, $H_{sf}$ and several
expressions of the mapping procedure. In $H_{GtJ}$ it has the only effect of
changing the sign of the nearest-neighbor hopping $t_{1}^{\prime }$. This
shifts in $(\pi ,\pi )$ the one-particle wave vectors. In particular,
restoring the original phases, the oxygen intensity (Eq. (\ref{14})) is
proportional to $\sin ^{2}(k_{x}/2)+\sin ^{2}(k_{y}/2)$.

\bibitem{bar}  F. Barriquand and G.A. Sawatzky, Phys. Rev. B {\bf 50}, 16649
(1994).

\bibitem{raman}  R.R.P. singh, P.A. Fleiry, K.B. Lyons and P.E. Sulewski,
Phys. Rev. Lett. {\bf 63}, 2736 (1989); G. Blumberg {\it et al.},Phys. Rev. B%
{\bf 53}, 11930( 1996).

\bibitem{note}  Orthogonal Zhang-Rice states can also be used in the mapping
procedure \cite{sim3,fei1,bel2,zha,ali}, but the numerical results \cite
{val,ero,tri} and an analytical comparison of both mappings \cite{ali}
indicates that for realistic or large O-O hopping $t_{pp}$, more accurate
results are obtained using non-orthogonal singlets, without having to
introduce higher order corrections \cite{fei1,tri}.

\bibitem{tri}  M.E. Simon and A.A. Aligia, Phys. Rev. B {\bf 52}, 7701
(1995).

\bibitem{rvb1}  C.D. Batista and A.A. Aligia, Physica C {\bf 261} 237 (1996).

\bibitem{rvb2}  C.D. Batista, L.O. Manuel, H.A. Ceccatto, and A.A. Aligia,
Europhys. Lett. {\bf 38}, 147 (1997); J. Low Temp. Phys. {\bf 105}, 591
(1996); F. Lema, C.D. Batista and A.A. Aligia, Physica C {\bf 259,} 287
(1996)., 5637 (1997).

\bibitem{shen}  Z-X. Shen, private communication.
\end{references}
\end{document}